# Fluctuation-Electromagnetic Interaction of Rotating Neutral Particle with the Surface: Relativistic Theory

A.A. Kyasov and G.V. Dedkov

Based on the fluctuation-electromagnetic theory, we have calculated the retarded force of attraction, frictional moment and heating rate of a neutral particle rotating near a polarizable surface. The particle and surface are characterized by arbitrary material properties and temperatures. The surface is assumed to be in thermal equilibrium with vacuum environment.

## 1. Introduction

Rotation of a body along with the uniform motion affects his coupling with the fluctuation electromagnetic field of other bodies. This results in the change of the van –der –Waals –Casimir forces (attractive and dissipative), and radiative (vacuum) heat exchange [1—3]. In a series of our papers [4,5] we have examined the impact of rotation [4] and a combined effect of rotation and uniform motion [5] on the fluctuation electromagnetic interaction in different configurations of the interacting subsystems. The obtained integral expressions depend on the direction of rotation, while subsystems are characterized by arbitrary dielectric and magnetic properties, and different temperatures. The aim of this paper is to consider the impact of rotation on the fluctuation electromagnetic interaction between a small polalizable particle and the surface (homogeneous half-space) with allowance for the retardation effects. The corresponding non-retarded limit was considered in [2,3] and all basic formulas in [2,3] follow from our general (retarded) formulas in this work as a partial case.

## 2. Theory

Consider a small spherical particle of radius $R$ with the electric and magnetic polarizabilities $\alpha_e(\omega), \alpha_m(\omega)$, and temperature $T_1$, rotating with the angular velocity $\mathbf{\Omega}=(0,0,\Omega)$ and located a distance $z_0$ apart from the surface (Fig.1)

Surface $z=0$ separates the vacuum half-space ($z>0$) from the half-space ($z<0$) filled by a continuous medium with dielectric and magnetic permittivities $\varepsilon(\omega), \mu(\omega)$, and temperature $T_2$. Assuming the conditions

$$R << z_0,\; R << \min\{c/\Omega, 2\pi\hbar c/k_B T_1, 2\pi\hbar c/k_B T_2\}$$

to be fulfilled, one can consider rotating particle as a point-like fluctuating electric and magnetic dipole. In addition, due to the relativistic problem statement, the value of $z_0$ has no upper restrictions.

According to the well-accepted method of calculation (for a review see [6]), with allowance for spontaneous and induced fluctuations $\mathbf{d}, \mathbf{m}, \mathbf{E}, \mathbf{B}$ in the system $\Sigma$, the initial expressions for calculating the force of attraction to the surface $F_z$, the rate of heating (cooling) of the particle $\dot{Q}$, and the frictional moment $M_z$ have the form

$$F_z = \left\langle \nabla_z (\mathbf{d}^{sp}\mathbf{E}^{in} + \mathbf{m}^{sp}\mathbf{B}^{in}) \right\rangle + \left\langle \nabla_z (\mathbf{d}^{in}\mathbf{E}^{sp} + \mathbf{m}^{in}\mathbf{B}^{sp}) \right\rangle \tag{1}$$

$$\dot{Q} = \left\langle (\dot{\mathbf{d}}^{sp}\mathbf{E}^{in} + \dot{\mathbf{m}}^{sp}\mathbf{B}^{in}) \right\rangle + \left\langle (\dot{\mathbf{d}}^{in}\mathbf{E}^{sp} + \dot{\mathbf{m}}^{in}\mathbf{B}^{sp}) \right\rangle \tag{2}$$

$$M_z = \left\langle \left[\mathbf{d}^{sp} \times \mathbf{E}^{in}\right]_z + \left[\mathbf{m}^{sp} \times \mathbf{B}^{in}\right]_z + \left[\mathbf{d}^{in} \times \mathbf{E}^{sp}\right]_z + \left[\mathbf{m}^{in} \times \mathbf{B}^{sp}\right]_z \right\rangle \tag{3}$$

where $\nabla_z = \partial/\partial z$, and the upper points above $Q, \mathbf{d}, \mathbf{m}$ denote the time derivatives.

In what follows, it should be taken into consideration that in the case of rotation around the $z$ axis, the fluctuation-dissipation relations for the spontaneous fluctuating electric and magnetic moments of the particle in the system $\Sigma$ of the surface take the form

$$\left\langle d_z^{sp}(\omega') d_z^{sp}(\omega) \right\rangle = 2\pi \hbar \delta(\omega + \omega') \alpha_e''(\omega) \coth(\hbar\omega/2k_B T_1) \tag{4}$$

$$\left\langle d_x^{sp}(\omega') d_x^{sp}(\omega) \right\rangle = \left\langle d_y^{sp}(\omega') d_y^{sp}(\omega) \right\rangle = \frac{1}{2} 2\pi \hbar \delta(\omega + \omega') \cdot$$
$$\cdot \left[\alpha_e''(\omega_+) \coth(\hbar\omega_+/2k_B T_1) + \alpha_e''(\omega_-) \coth(\hbar\omega_-/2k_B T_1)\right] \tag{5}$$

$$\left\langle d_x^{sp}(\omega') d_y^{sp}(\omega) \right\rangle = -\left\langle d_y^{sp}(\omega') d_x^{sp}(\omega) \right\rangle = -\frac{i}{2} 2\pi \hbar \delta(\omega + \omega') \cdot$$
$$\cdot \left[\alpha_e''(\omega_+) \coth(\hbar\omega_+/2k_B T_1) - \alpha_e''(\omega_-) \coth(\hbar\omega_-/2k_B T_1)\right] \tag{6}$$

$$\left\langle m_z^{sp}(\omega') m_z^{sp}(\omega) \right\rangle = 2\pi \hbar \delta(\omega + \omega') \alpha_m''(\omega) \coth(\hbar\omega/2k_B T_1) \tag{7}$$

$$\left\langle m_x^{sp}(\omega')m_x^{sp}(\omega)\right\rangle = \left\langle m_y^{sp}(\omega')m_y^{sp}(\omega)\right\rangle = \frac{1}{2}2\pi\hbar\delta(\omega+\omega')\cdot$$
$$\cdot\left[\alpha_m''(\omega_+)\coth(\hbar\omega_+/2k_BT_1) + \alpha_m''(\omega_-)\coth(\hbar\omega_-/2k_BT_1)\right]\tag{8}$$

$$\left\langle m_x^{sp}(\omega')m_y^{sp}(\omega)\right\rangle = -\left\langle m_y^{sp}(\omega')m_x^{sp}(\omega)\right\rangle = -\frac{i}{2}2\pi\hbar\delta(\omega+\omega')\cdot$$
$$\cdot\left[\alpha_m''(\omega_+)\coth(\hbar\omega_+/2k_BT_1) - \alpha_m''(\omega_-)\coth(\hbar\omega_-/2k_BT_1)\right]\tag{9}$$

where $\omega_\pm = \omega \pm \Omega$.

In addition, relations between the induced dipole and magnetic moments of rotating particle and spontaneous fluctuating electromagnetic field of the surface in the reference frame $\Sigma$ are given by

$$d_z^{in}(t) = \int\frac{d\omega d^2k}{(2\pi)^3}\alpha_e(\omega)E_z^{sp}(\omega,\mathbf{k})\exp(-i\omega t)\tag{10}$$

$$d_x^{in}(t) = \int\frac{d\omega d^2k}{(2\pi)^3}\frac{1}{2}\exp(-i\omega t)\begin{bmatrix}\alpha_e(\omega_+)\left(E_x^{sp}(\omega,\mathbf{k})+iE_y^{sp}(\omega,\mathbf{k})\right)+\\+\alpha_e(\omega_-)\left(E_x^{sp}(\omega,\mathbf{k})-iE_y^{sp}(\omega,\mathbf{k})\right)\end{bmatrix}\tag{11}$$

$$d_y^{in}(t) = \int\frac{d\omega d^2k}{(2\pi)^3}\frac{1}{2}\exp(-i\omega t)\begin{bmatrix}\alpha_e(\omega_+)\left(E_y^{sp}(\omega,\mathbf{k})-iE_x^{sp}(\omega,\mathbf{k})\right)+\\+\alpha_e(\omega_-)\left(E_y^{sp}(\omega,\mathbf{k})+iE_x^{sp}(\omega,\mathbf{k})\right)\end{bmatrix}\tag{12}$$

$$m_z^{in}(t) = \int\frac{d\omega d^2k}{(2\pi)^3}\alpha_m(\omega)B_z^{sp}(\omega,\mathbf{k})\exp(-i\omega t)\tag{13}$$

$$m_x^{in}(t) = \int\frac{d\omega d^2k}{(2\pi)^3}\frac{1}{2}\exp(-i\omega t)\begin{bmatrix}\alpha_m(\omega_+)\left(B_x^{sp}(\omega,\mathbf{k})+iB_y^{sp}(\omega,\mathbf{k})\right)+\\+\alpha_m(\omega_-)\left(B_x^{sp}(\omega,\mathbf{k})-iB_y^{sp}(\omega,\mathbf{k})\right)\end{bmatrix}\tag{14}$$

$$m_y^{in}(t) = \int\frac{d\omega d^2k}{(2\pi)^3}\frac{1}{2}\exp(-i\omega t)\begin{bmatrix}\alpha_m(\omega_+)\left(B_y^{sp}(\omega,\mathbf{k})-iB_x^{sp}(\omega,\mathbf{k})\right)+\\+\alpha_m(\omega_-)\left(B_y^{sp}(\omega,\mathbf{k})+iB_x^{sp}(\omega,\mathbf{k})\right)\end{bmatrix}\tag{15}$$

It is worth noting that all Fourier-transforms $\mathbf{E}^{sp}(\omega,\mathbf{k}), \mathbf{B}^{sp}(\omega,\mathbf{k})$ are taken at the point $z_0$.

### 3. Results

As a result of the following standard calculations, with allowance for (1)—(15) we obtain (one- and double-primed quantities denote real and imaginary components of the functions)

$$F_z = -\frac{\hbar}{4\pi^2} \int_{-\infty}^{+\infty} d\omega \int_{-\infty}^{+\infty} dk_x \int_{-\infty}^{+\infty} dk_y \cdot$$

$$\left\{ \begin{array}{l} \alpha_e''(\omega_+) \operatorname{Re}\left( \exp(-2q_0 z_0) \left[ \left( k^2 - \frac{\omega^2}{c^2} \right) \Delta_e(\omega) + \frac{\omega^2}{c^2} \Delta_m(\omega) \right] \right) \coth\left( \frac{\hbar \omega_+}{2k_B T_1} \right) + \\ + \alpha_e'(\omega_+) \operatorname{Im}\left( \exp(-2q_0 z_0) \left[ \left( k^2 - \frac{\omega^2}{c^2} \right) \Delta_e(\omega) + \frac{\omega^2}{c^2} \Delta_m(\omega) \right] \right) \coth\left( \frac{\hbar \omega}{2k_B T_2} \right) + \\ + \alpha_e''(\omega) \operatorname{Re}\left( \exp(-2q_0 z_0) k^2 \Delta_e(\omega) \right) \coth\left( \frac{\hbar \omega}{2k_B T_1} \right) + \\ + \alpha_e'(\omega) \operatorname{Im}\left( \exp(-2q_0 z_0) k^2 \Delta_e(\omega) \right) \coth\left( \frac{\hbar \omega}{2k_B T_2} \right) + (e \leftrightarrow m) \end{array} \right\} \qquad (16)$$

$$\dot{Q} = \frac{\hbar}{4\pi^2} \int_{-\infty}^{+\infty} d\omega \int_{-\infty}^{+\infty} dk_x \int_{-\infty}^{+\infty} dk_y \cdot \omega \cdot$$

$$\left\{ \begin{array}{l} \alpha_e''(\omega_+) \operatorname{Im}\left( \frac{\exp(-2q_0 z_0)}{q_0} \left[ \left( k^2 - \frac{\omega^2}{c^2} \right) \Delta_e(\omega) + \frac{\omega^2}{c^2} \Delta_m(\omega) \right] \right) \cdot \\ \cdot \left[ \coth\left( \frac{\hbar \omega}{2k_B T_2} \right) - \coth\left( \frac{\hbar \omega_+}{2k_B T_1} \right) \right] + \\ + \alpha_e''(\omega) \operatorname{Im}\left( \frac{\exp(-2q_0 z_0)}{q_0} k^2 \Delta_e(\omega) \right) \cdot \\ \cdot \left[ \coth\left( \frac{\hbar \omega}{2k_B T_2} \right) - \coth\left( \frac{\hbar \omega}{2k_B T_1} \right) \right] + (e \leftrightarrow m) \end{array} \right\} \qquad (17)$$

$$M_z = -\frac{\hbar}{2\pi^2} \int_{-\infty}^{+\infty} d\omega \int_{-\infty}^{+\infty} dk_x \int_{-\infty}^{+\infty} dk_y \cdot$$

$$\cdot \left\{ \begin{array}{l} \alpha_e''(\omega_+) \operatorname{Im}\left( \frac{\exp(-2q_0 z_0)}{q_0} \left[ \left( k^2 - \frac{\omega^2}{c^2} \right) \Delta_e(\omega) + \frac{\omega^2}{c^2} \Delta_m(\omega) \right] \right) \cdot \\ \cdot \left[ \coth\left( \frac{\hbar \omega}{2k_B T_2} \right) - \coth\left( \frac{\hbar \omega_+}{2k_B T_1} \right) \right] + (e \leftrightarrow m) \end{array} \right\} \qquad (18)$$

where $q_0 = \sqrt{k^2 - \omega^2/c^2}$, $q = \sqrt{k^2 - (\omega^2/c^2)\varepsilon(\omega)\mu(\omega)}$, $k^2 = k_x^2 + k_y^2$,

$\Delta_e(\omega) = \dfrac{q_0 \varepsilon(\omega) - q}{q_0 \varepsilon(\omega) + q}$, $\Delta_e(\omega) = \dfrac{q_0 \mu(\omega) - q}{q_0 \mu(\omega) + q}$. Moreover, the bracketed terms $(e \leftrightarrow m)$ in Eqs. (16)—(18) are given by the explicitly written terms with the replacements $\alpha_e(\omega) \to \alpha_m(\omega)$, $\Delta_e(\omega) \to \Delta_m(\omega)$, $\Delta_m(\omega) \to \Delta_e(\omega)$.

### 4. Rotation axis parallel to the surface

The case where the rotation axis is parallel to the surface is quite similar to the case $\mathbf{\Omega} \parallel z$. Obviously, the configurations $\mathbf{\Omega} = (\Omega,0,0)$ and $\mathbf{\Omega} = (0,\Omega,0)$ corresponding to $\mathbf{\Omega} \parallel x$ and $\mathbf{\Omega} \parallel y$. are equivalent. Fluctuation-dissipation relations (4)-(9) and Eqs. (10)—(15) should be rewritten using a cyclic permutation $x \to y \to z \to x$. Then, in the case $\mathbf{\Omega} \parallel x$, equations analogous to (16)—(18) take the form

$$F_z = -\dfrac{\hbar}{4\pi^2} \int\limits_{-\infty}^{+\infty} d\omega \int\limits_{-\infty}^{+\infty} dk_x \int\limits_{-\infty}^{+\infty} dk_y \cdot$$

$$\left\{ \begin{array}{l} \alpha_e''(\omega_+) \operatorname{Re}\left( \exp(-2q_0 z_0) \left[ \left( k_y^2 + k^2 - \dfrac{\omega^2}{2c^2} \right) \Delta_e(\omega) + \dfrac{\omega^2}{2c^2} \Delta_m(\omega) \right] \right) \coth\left( \dfrac{\hbar \omega_+}{2k_B T_1} \right) + \\[6pt] + \alpha_e'(\omega_+) \operatorname{Im}\left( \exp(-2q_0 z_0) \left[ \left( k_y^2 + k^2 - \dfrac{\omega^2}{2c^2} \right) \Delta_e(\omega) + \dfrac{\omega^2}{2c^2} \Delta_m(\omega) \right] \right) \coth\left( \dfrac{\hbar \omega}{2k_B T_2} \right) + \\[6pt] + \alpha_e''(\omega) \operatorname{Re}\left( \exp(-2q_0 z_0) \left[ \left( k_x^2 - \dfrac{\omega^2}{2c^2} \right) \Delta_e(\omega) + \dfrac{\omega^2}{2c^2} \Delta_m(\omega) \right] \right) \coth\left( \dfrac{\hbar \omega}{2k_B T_1} \right) + \\[6pt] + \alpha_e'(\omega) \operatorname{Im}\left( \exp(-2q_0 z_0) \left[ \left( k_x^2 - \dfrac{\omega^2}{2c^2} \right) \Delta_e(\omega) + \dfrac{\omega^2}{2c^2} \Delta_m(\omega) \right] \right) \coth\left( \dfrac{\hbar \omega}{2k_B T_2} \right) + (e \leftrightarrow m) \end{array} \right\} \quad (19)$$

$$\dot{Q} = \dfrac{\hbar}{4\pi^2} \int\limits_{-\infty}^{+\infty} d\omega \int\limits_{-\infty}^{+\infty} dk_x \int\limits_{-\infty}^{+\infty} dk_y \cdot \omega \cdot$$

$$\cdot \left\{ \begin{array}{l} \alpha_e''(\omega_+) \operatorname{Im}\left( \dfrac{\exp(-2q_0 z_0)}{q_0} \left[ \left( k_y^2 + k^2 - \dfrac{\omega^2}{2c^2} \right) \Delta_e(\omega) + \dfrac{\omega^2}{2c^2} \Delta_m(\omega) \right] \right) \cdot \\[6pt] \cdot \left[ \coth\left( \dfrac{\hbar \omega}{2k_B T_2} \right) - \coth\left( \dfrac{\hbar \omega_+}{2k_B T_1} \right) \right] + \\[6pt] \alpha_e''(\omega) \operatorname{Im}\left( \dfrac{\exp(-2q_0 z_0)}{q_0} \left[ \left( k_x^2 - \dfrac{\omega^2}{2c^2} \right) \Delta_e(\omega) + \dfrac{\omega^2}{2c^2} \Delta_m(\omega) \right] \right) \cdot \\[6pt] \cdot \left[ \coth\left( \dfrac{\hbar \omega}{2k_B T_2} \right) - \coth\left( \dfrac{\hbar \omega}{2k_B T_1} \right) \right] + (e \leftrightarrow m) \end{array} \right\} \quad (20)$$

$$M_x = -\frac{\hbar}{2\pi^2} \int_{-\infty}^{+\infty} d\omega \int_{-\infty}^{+\infty} dk_x \int_{-\infty}^{+\infty} dk_y \cdot$$

$$\cdot \left\{ \alpha_e''(\omega_+) \mathrm{Im}\left( \frac{\exp(-2q_0 z_0)}{q_0} \left[ \left( k_y^2 + k^2 - \frac{\omega^2}{2c^2} \right) \Delta_e(\omega) + \frac{\omega^2}{2c^2} \Delta_m(\omega) \right] \right) \cdot \right. \quad (21)$$

$$\left. \cdot \left[ \coth\left(\frac{\hbar\omega}{2k_B T_2}\right) - \coth\left(\frac{\hbar\omega_+}{2k_B T_1}\right) \right] + (e \leftrightarrow m) \right\}$$

In the case $\Omega \parallel y$, Eqs. (19)—(21) are the same with the replacements $x \to y$, $k_y^2 \leftrightarrow k_x^2$. Moreover, it is easy to verify that at $\Omega = 0$ Eqs. (16), (19) and Eqs. (17), (20) are transformed into one another and describe the static Casimir-Polder force and the particle-surface radiative heat exchange when the surface is in thermal equilibrium with the vacuum environment but out of equilibrium with the particle [6]. On the other hand, when neglecting retardation effects (in the limit $c \to \infty$), Eqs.(16)—(21) reduce to the corresponding nonretarded results obtained in [3,4].

## 5. Ideally conducting surface and particle

Let us consider the simplest case $T_1 = T_2 = T = 0$, assuming the interacting subsystems to be ideally conducting. In this case we must take

$$\Delta_e(\omega) = 1, \Delta_m(\omega) = -1, \alpha_e(\omega) = R^3, \alpha_m(\omega) = -R^3/2, \coth(\hbar\omega/2k_B T) = sign(\omega), \quad (22)$$
$$\coth(\hbar\omega_+/2k_B T) = sign(\omega + \Omega)$$

Substituting (22) into (16) and carrying out integrations yields the well-known result [6,7]

$$F_z = -\frac{9}{4\pi} \frac{\hbar c R^3}{z_0^5} \quad (23)$$

corresponding to an ideally conducting immovable particle and ideally conducting half-space. The same result is obtained under the conditions of finite conductivity in the limiting case of strong retardation, where the particle-surface distance is much greater than the characteristic wave-length in their absorption spectra [8]. The difference in the numerical coefficients in [6,7] and [8] is due to the lack of the effect of magnetic polarization in [8].

Therefore, within the accepted assumptions, rotation of the particle does not alter the Casimir-Polder force. This result can also be formulated as follows: provided that imaginary parts of the particle polarizabilities are equal to zero, there is no coupling between the particle rotation and zero oscillations of the surface electromagnetic field. Moreover, at $T_1 = T_2 = T = 0$, for ideally conducting particle and surface, frictional moment and heating rate of the particle disappear.

Contrary to that, the effect of rotation will be appreciable in the case of real material properties of the particle (surface) or both of them, as well as in the case of different temperatures.

## 6. Conclusions

For the first time, we have generalized the recently obtained theoretical expressions for the fluctuation electromagnetic attraction force, frictional moment and the rate of radiative heat exchange between a rotating particle and the surface with allowance for the effect of retardation. The earlier formulas in the evanescent-field limit follow from the general results as a partial case. We have shown that in the case of ideally conducting particle and surface, angular rotation of the particle does not influence the Casimir-Polder force at zero temperature of the system, while the frictional moment and heating rate (at $T_1 = T_2 = T = 0$) are equal to zero.

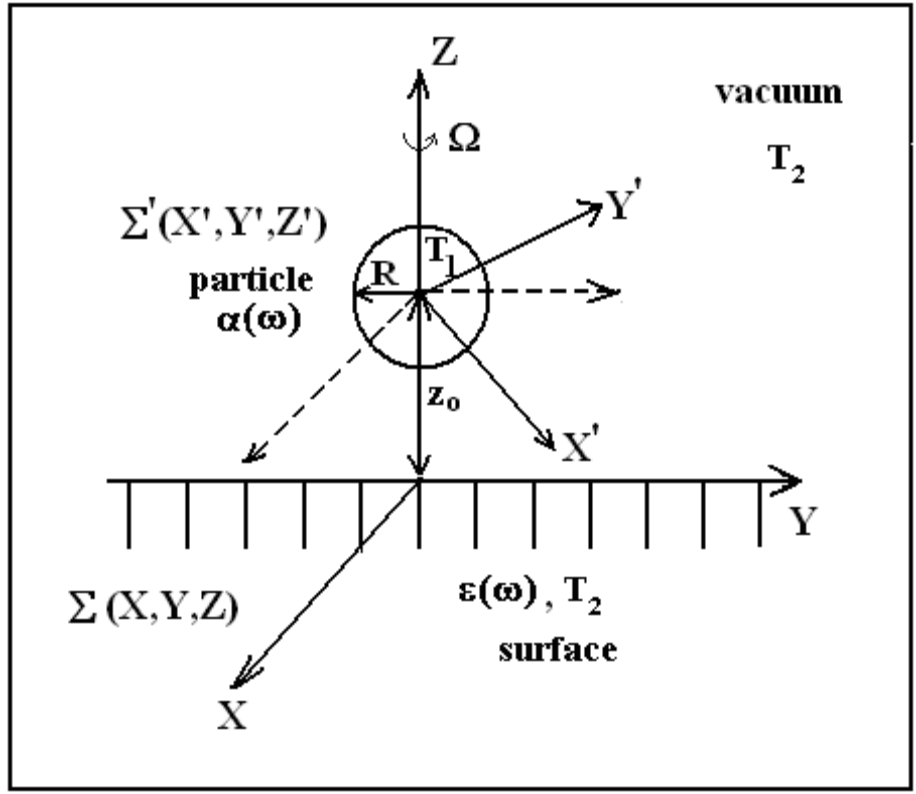

Fig. 1 Geometrical configuration and coordinate systems used.